\documentclass[floatfix,twocolumn,prl,longbibliography]{revtex4-1}
\usepackage{graphicx}
\usepackage{dcolumn}% Align table columns on decimal point
\usepackage{bm}% bold math
\usepackage{amsmath}
\usepackage{times}
\usepackage{color}
\usepackage[breaklinks=true,colorlinks,citecolor=blue,linkcolor=blue,urlcolor=blue]{hyperref}
\usepackage{gensymb}
\makeatletter

\newcommand{\Rmnum}[1]{\expandafter\@slowromancap\romannumeral #1@}
\makeatother

\begin{document}
\title{In-plane anomalous and third-harmonic Hall response in an easy-plane trigonal magnet}
\author{Arnab Das}
\affiliation{Department of Physics, Indian Institute of Technology Kanpur, Kanpur 208016, India}

\author{Soumik Mukhopadhyay}
\email{soumikm@iitk.ac.in}
\affiliation{Department of Physics, Indian Institute of Technology Kanpur, Kanpur 208016, India}  
    
\begin{abstract}
    We report the emergence of an in-plane anomalous Hall effect (IPAHE) and a pronounced third-harmonic Hall response in an easy-plane trigonal magnetic system achieved via tuning of magnetic anisotropy. Angle-dependent Hall measurements reveal a clear departure from conventional sinusoidal behavior, revealing a strong sin($\mathrm{3\theta}$) component at elevated fields, which increases as the magnetic anisotropy is reduced, accompanied by a finite in-plane Hall signal. Consistently, the observed nonlinear Hall signal exhibits a cubic field dependence, indicative of a leading third-order contribution to the transverse conductivity. The enhancement of the third harmonic reflects the evolution toward coherent in-plane magnetization and symmetry-governed nonlinear transport. These observations support an intrinsic, symmetry-controlled transport mechanism enabled by reduced magnetic anisotropy and establish a direct link between anisotropy, crystal symmetry, and higher-order Hall response.
\end{abstract}

\maketitle

{\it Introduction:---}
The conventional Hall effect, one of the most important phenomena in condensed matter physics, is the generation of a voltage transverse to an electric current under an out-of-plane magnetic field \cite{Hall1879, Hall1881, hurd2012hall, PhysRevLett.133.236602}. A much more interesting phenomenon was observed in ferromagnetic (FM) materials, where the Hall voltage is observed in the absence of a magnetic field and was termed the anomalous Hall effect (AHE) \cite{RevModPhys.82.1539}. The conventional AHE usually occurs when the magnetic field/magnetization is perpendicular to the measurement plane. Recently, a planar Hall effect (PHE), in which the magnetic field/magnetization is in-plane, has gained great importance \cite{Zhang2011,Liu2013,Ren2016,Zhong2017,Liu2018QAH,Sun2020,Zyuzin2020,Battilomo2021,Cullen2021,Tan2021,Sun2022,Cao2023,Li2023PHE,Wang2024,Xiang2024}. Beyond the conventional AHE, where the transverse response is primarily driven by the out-of-plane magnetization, magnetic materials can also exhibit an in-plane anomalous Hall effect (IPAHE) \cite{PhysRevB.109.155408,PhysRevB.111.054412,
https://doi.org/10.1002/adma.202502624,PhysRevLett.133.236602}. In this case, Onsager symmetry permits a transverse Hall response even when the magnetization or magnetic field lies within the plane defined by the electric field and Hall current \cite{landau2013electrodynamics}. The IPAHE emerges from symmetry-allowed transverse conductivity generated by in-plane magnetization in the presence of strong spin–orbit coupling and reduced crystalline symmetry. Such behavior is generally forbidden in high-symmetry systems, but can become symmetry allowed when specific rotational or mirror symmetries are broken \cite{PhysRevResearch.5.023138}. The IPAHE is a field-odd effect and is prohibited in systems possessing C$_2$, C$_4$, or C$_6$ rotational symmetry along the out of plane axis \cite{PhysRevLett.133.236602}. Only systems possessing C$_1$ or C$_3$ rotational symmetry can give rise to a non-zero in-plane anomalous Hall voltage. The IPAHE also differs from the PHE, which happens to be a field-even effect arising from the anisotropy in magnetoresistance (MR) \cite{PhysRevResearch.5.023138}. 

In trigonal and noncentrosymmetric magnetic systems, the absence of specific mirror symmetries permits unconventional transverse conductivity tensors and enables higher-order Hall responses beyond the linear regime. In trigonal systems possessing threefold (C$_3$) rotational symmetry, the in-plane magnetic free energy can acquire higher-order angular terms of the form $
F(\phi) \sim -\mathbf{M}\cdot\mathbf{B} + K_3\cos(3\phi),
$ where $\phi$ denotes the in-plane magnetization angle and $K_3$ represents the trigonal anisotropy constant. Such symmetry-allowed anisotropic terms naturally generate higher-order angular harmonics in the Hall response and can give rise to nonlinear contributions in angle-dependent Hall measurements.

It has been suggested that in trigonal magnetic systems belonging to the $\bar{3}m$ symmetry class, the symmetry requirements for a non-zero IPAHE signal are satisfied  \cite{PhysRevResearch.5.023138}. In this perspective, the layered ferrimagnet Mn$_3$Si$_2$Te$_6$ (MST) which crystallizes in the trigonal space group $P\bar{3}1c$ and exhibits an easy-plane magnetic ground state, becomes an attractive platform for investigating the interplay between magnetic anisotropy and crystal symmetry. \textcolor{blue}{Recent theoretical works identify MST as a candidate material for
spontaneous IPAHE due to the threefold rotational symmetry, which can be switched by changing the magnetization orientation with external
magnetic fields \cite{PhysRevB.109.155153, luo2026unifiedsymmetryframeworkinplane}.} Previous studies have also demonstrated that chemical substitution provides an effective route for tuning its magnetic anisotropy and in-plane magnetic response \cite{PhysRevB.111.174419,Zhang2022ChiralOrbitalCurrents}.

Here, we report the emergence of an IPAHE and pronounced third-harmonic Hall response in nanoflake Mn$_3$Si$_2$(Te$_{1-x}$Se$_x$)$_6$ (MSTS$x$) through Se substitution and anisotropy tuning. Angle-dependent Hall measurements reveal a strong third-harmonic contribution at elevated magnetic fields, which increases systematically as magnetic anisotropy is reduced, alongside a finite in-plane Hall signal. Harmonic decomposition establishes a direct correlation between the enhanced IPAHE and third-harmonic response, highlighting the increasing role of trigonal crystal symmetry in transverse transport characteristics.

\begin{figure}
  \includegraphics[width=\linewidth]{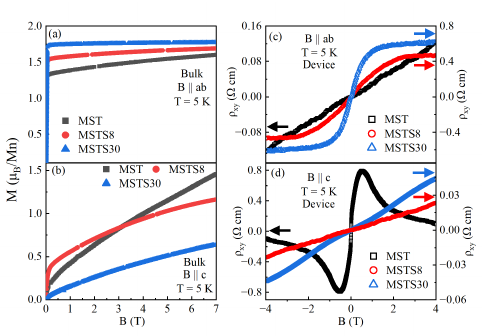}
  \caption{Isothermal magnetization M(B) of Mn$_3$Si$_2$(Te$_{1-x}$Se$_x$)$_6$ single crystals for different concentrations of selenium with (a) B $\|$ ab and (b) B $\|$ c directions, respectively. Hall resistivity curve at 5 K for devices of Mn$_3$Si$_2$(Te$_{1-x}$Se$_x$)$_6$ for different concentrations of selenium with (c) B $\|$ ab and (d) B $\|$ c directions, respectively. The arrows indicate the scale, followed by the curves.}
  \label{fig1}
\end{figure}
{\it Experimental Methods:---}
Single crystals of MST and MSTS$x$ are prepared using the chemical vapour transport (CVT) method, where iodine (I$_2$) is used as the transport agent (see Appendix\hyperref[appendixA]{A} for details on sample preparation). The crystals are typically of 1-2 mm size. Energy dispersive spectroscopy (EDS) measurements are performed to confirm the elemental composition of the grown crystals. X-ray diffraction (XRD) measurements are performed using a PANalytical Empyrean diffractometer. Both powder XRD and single crystal XRD measurements are carried out to confirm the purity and crystallinity of the grown crystals. The powder XRD data of both MST and MSTS$x$ are profile-fitted using P$\Bar{3}$1c space group symmetry using FullProf$\_$Suite software. The temperature dependence of magnetization and the isothermal magnetization measurements of the bulk single crystals are carried out using a Quantum Design Physical Properties Measurement System (PPMS). Devices with Hall bar geometry are fabricated using the dry-transfer method. Standard Hall bar electrodes of Ti/Au (15/40 nm) are pre-patterned on a 300 nm oxidized SiO$_2$/Si substrate via electron beam lithography (EBL), followed by e-beam deposition and a lift-off process. Scotch tape is used for the mechanical exfoliation of the bulk crystals and to transfer them onto polydimethylsiloxane (PDMS) stamps. Flakes with suitable thickness and shape are selected and transferred onto the pre-patterned electrodes using an XYZ micromanipulator combined with an optical microscope. The entire transfer process is performed in a glove box with N$_2$ atmosphere to prevent any exposure to oxygen or moisture. For magneto-transport measurements, we use a variable temperature insert (VTI) cryostat manufactured by CRYOGENIC Inc., UK. The raw MR and Hall curves are symmetrized and antisymmetrized, respectively, to eliminate the effects of misalignment of the electrode.

{\it Results and Discussion:---}
The single-crystal XRD data of the bulk single crystal reveals the (001) orientation of the crystal plane (see Appendix\hyperref[appendixA]{A} for details). The powder XRD spectra (see SM-Fig.~1(e)-(g) of \cite{PhysRevB.111.174419} for details) suggest that the single crystals grown are of high purity. Additionally, the EDS spectra (see Appendix\hyperref[appendixA]{A} for details) shows prominent peaks of Mn, Si, Te and Se, confirming optimum stoichiometry. 

The isothermal magnetization M(B) of bulk MSTS$x$ single crystals for different values of $x$ is shown in Fig.~\ref{fig1}(a),(b) for B $\|$ ab and B $\|$ c directions respectively. The ab-plane remains the easy plane even after increased Se substitution. From Fig.~\ref{fig1}(a), it is visible that with increasing concentration of Se, i.e., with increasing value of $x$, the magnetization, with field applied along the ab-plane, saturates faster. The saturation magnetization also increases with increasing value of $x$. However, the scenario reverses when the applied field is along the magnetic hard axis, i.e., the c-axis. Figure~\ref{fig1}(b) shows that when B $\|$ c, the magnetization remains unsaturated upon Se substitution, and also with increasing value of $x$, the high-field magnetization along the c-axis decreases, in agreement with prior reports \cite{PhysRevB.111.174419}. 

Figure~\ref{fig1}(c),(d) shows the Hall resistivity $\mathrm{\rho_{xy}}$ for MSTS$x$ for different values of $x$ at 5 K along B $\|$ ab and B $\|$ c directions, respectively. For $x$ = 0 (i.e., for MST), the Hall resistivity along B $\|$ c shows the conventional topological Hall response as reported earlier \cite{Das2025ChiralOC,PhysRevB.108.125103,PhysRevB.103.L161105,Zhang2024CurrentSensitiveHall}. As we increase $x$, the topological Hall signal disappears, and we get a standard linear $\mathrm{\rho_{xy}}$. However, the Hall response along B $\|$ ab shows some interesting behavior as is visible in Fig.~\ref{fig1}(c). The $\mathrm{\rho_{xy}}$ shows a linear behaviour for MST device, but as we increase the value of $x$, nonlinear contributions start to show up in $\mathrm{\rho_{xy}}$. For both MSTS8 and MSTS30, $\mathrm{\rho_{xy}}$ shows anomalous behavior, which is seen to increase with increasing value of $x$. The anomalous Hall effect starts to appear for MSTS8 but strengthens more for MSTS30, showing a pronounced IPAHE.
\begin{figure}
  \includegraphics[width=0.8\linewidth]{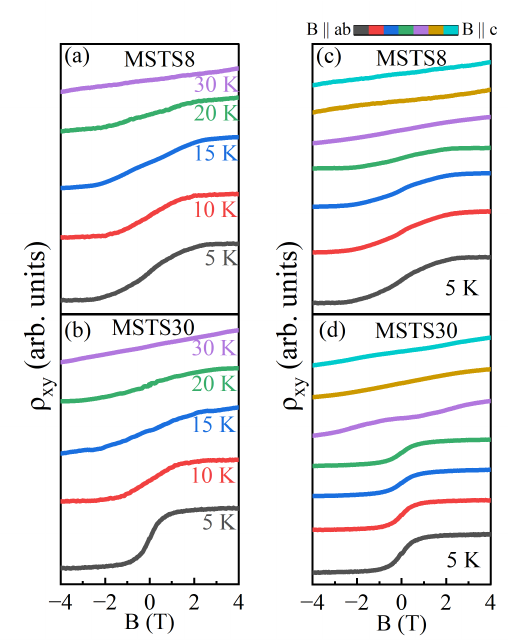}
  \caption{The field dependence of Hall resistivity ($\mathrm{\rho_{xy}}$) at different temperatures for (a) MSTS8 and (b) MSTS30 nanoflake device, respectively, along B $\|$ ab directions. The field dependence of Hall resistivity ($\mathrm{\rho_{xy}}$) at different angles for (c) MSTS8 and (d) MSTS30 nanoflake device, respectively, at 5 K.}
  \label{fig2}
\end{figure}
\begin{figure*}[htp]
  \includegraphics[width=\linewidth]{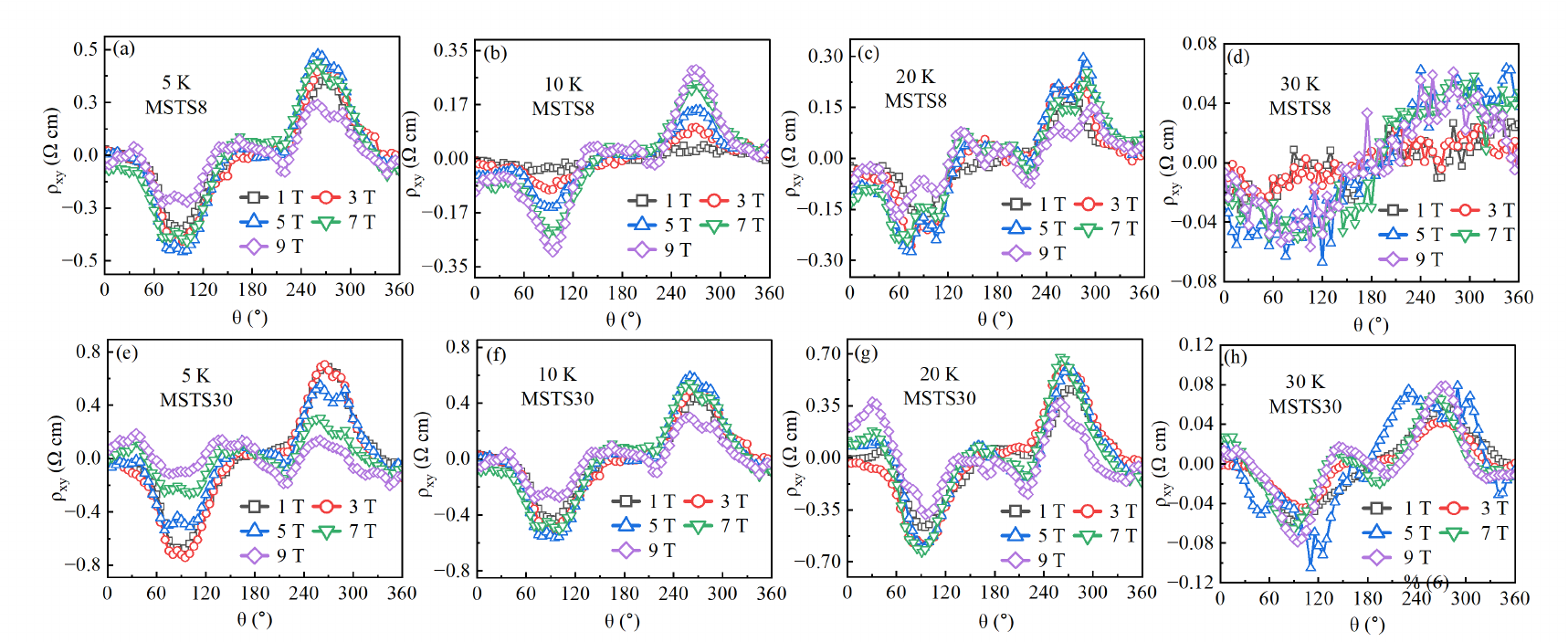}
  \caption{Angular dependence of the Hall resistivity ($\mathrm{\rho_{xy}}$) measured at different magnetic fields for MSTS8 [(a)–(d)] and MSTS30 [(e)–(h)] for different temperatures. The magnetic field was rotated in the plane containing the crystallographic c-axis and the current direction. Here, $\mathrm{\theta = 0^\circ}$ corresponds to B $\|$ c and $\mathrm{\theta = 90^\circ}$ corresponds to B $\|$ ab.}
  \label{fig3}
\end{figure*}

We further measured the Hall resistivity at different temperatures for both MSTS8 and MSTS30 devices, along the B $\|$ ab direction (see Appendix\hyperref[appendixB]{B} for details on extraction of the anomalous Hall contribution). The IPAHE is seen to weaken and disappear with increasing temperatures for both cases, as is evident from Fig.~\ref{fig2}(a),(b) respectively. However, both samples exhibit a finite Hall response despite the absence of an out-of-plane magnetic field component, which increases systematically at lower temperatures and becomes more pronounced with increasing Se concentration. In particular, MSTS30 exhibits a significantly enhanced nonlinear Hall response compared to MSTS8, consistent with the reduction of magnetic anisotropy and strengthening of easy-plane magnetic behavior upon Se substitution, as is visible in Fig.~\ref{fig1}(a). The reversible field dependence with negligible hysteresis further supports coherent in-plane magnetization reversal, characteristic of easy-plane magnetic systems. 

Figure~\ref{fig2}(c),(d) show the field dependence of Hall resistivity at 5 K for different field orientations varying from B $\|$ ab to B $\|$ c in MSTS8 and MSTS30, respectively. The black curve corresponds to the in-plane field configuration (B $\|$ ab), while the cyan curve represents the out-of-plane configuration (B $\|$ c); the intermediate curves correspond to field orientations between these two limits. A systematic evolution of the Hall response is observed as the magnetic field is rotated from the in-plane to out-of-plane direction, demonstrating a crossover between two transport regimes. For B $\|$ ab, a finite Hall signal is observed despite the absence of an out-of-plane magnetic field component, providing evidence for an in-plane anomalous Hall effect. In this configuration, the transverse response is dominated by the symmetry-allowed in-plane anomalous Hall channel originating predominantly from the coupling between the in-plane magnetization and the low-symmetry trigonal crystal structure. As the magnetic field is gradually rotated away from the ab-plane, an out-of-plane field component develops, leading to the progressive emergence of the conventional linear Hall response contribution, progressively masking the in-plane response. The absence of abrupt changes during rotation indicates that this crossover arises from continuous magnetization reorientation rather than a magnetic phase transition. This gradual redistribution highlights the competition between anisotropy-controlled in-plane transport and the conventional out-of-plane Hall effect.

The enhancement of the IPAHE with increasing Se concentration indicates that anisotropy tuning plays a central role in the emergence of unconventional transverse transport. Se substitution weakens the effective magnetic anisotropy and enhances the in-plane magnetic response \cite{PhysRevB.111.174419, Zhang2022ChiralOrbitalCurrents}. As a consequence, the magnetization becomes increasingly susceptible to the underlying trigonal crystal potential, strengthening the coupling between magnetic order and crystal symmetry. The stronger IPAHE observed in MSTS30 therefore suggests that reduced anisotropy promotes symmetry-allowed non-linear transverse transport channels that are otherwise weak or absent in the unsubstituted sample.

To examine the symmetry of the Hall response, we performed angle-dependent Hall measurements by rotating the magnetic field from B $\|$ c to B $\|$ ab while maintaining a constant field magnitude. For a conventional Hall response, the angular dependence is expected to be described by a single first-harmonic component arising from the projection of the magnetic field or magnetization. Any deviation from this behavior, therefore, provides direct evidence for additional symmetry-governed transport channels.
\begin{figure}
  \includegraphics[width=\linewidth]{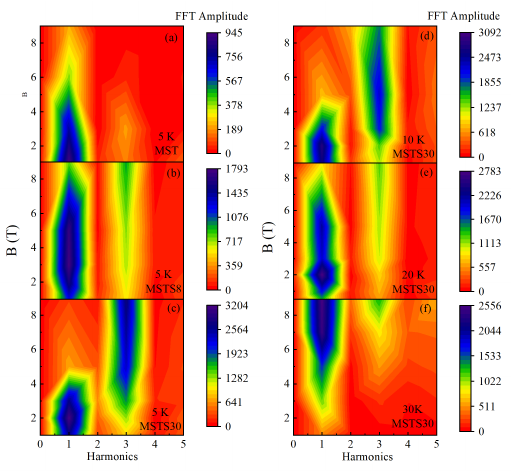}
  \caption{(a) - (c) Fast Fourier transform (FFT) analysis of the angular Hall resistivity $\mathrm{\rho_{xy}(\theta)}$ for MST, MSTS8 and MSTS30 at 5 K, respectively. (d) - (f) Fast Fourier transform (FFT) analysis of the angular Hall resistivity $\mathrm{\rho_{xy}(\theta)}$ for MSTS30 at 10, 20 and 30 K, respectively. The color maps show the FFT amplitudes of the harmonic components extracted from the angular Hall data as a function of harmonic order and magnetic field.}
  \label{fig4}
\end{figure}

Figure~\ref{fig3} reveals that the angular Hall response evolves from an approximately first-harmonic form at high temperatures and low magnetic fields to a strongly modulated profile at low temperatures and high fields, with the effect most pronounced in MSTS30. Notably, the recovery of a nearly sinusoidal angular dependence at 30 K coincides with the disappearance of the IPAHE, establishing a direct connection between the in-plane Hall channel and the higher-order harmonic response. To quantitatively analyze this behavior, the angular Hall resistivity was decomposed into harmonic components (see Appendix\hyperref[appendixC]{C} for more details on harmonic decomposition):
\begin{eqnarray}
    \mathrm{\rho_{xy}(\theta)}=\mathrm{A_1\sin(\theta)}+\mathrm{A_3\sin(3\theta)}+\mathrm{A_5\sin(5\theta)}+\cdots .
\end{eqnarray}
Remarkably, while the first harmonic component (A$_1$) remains dominant in MSTS8, the third-harmonic component (A$_3$) becomes pronounced in MSTS30 and exceeds $A_1$ at low temperatures and high fields. The appearance of the $\mathrm{\sin(3\theta)}$ term reflects the threefold rotational symmetry of the trigonal crystal structure and indicates that the Hall response can no longer be described solely by a conventional first-harmonic angular dependence.

To quantify the evolution of the angular Hall response, the harmonic amplitudes extracted from the Fast Fourier transform (FFT) analysis are summarized in Fig.~\ref{fig4} as a function of harmonic order and magnetic field, in a similar manner as reported in \cite{k5cj-2wd1}. Figure~\ref{fig4}(a)-(c) shows the FFT analysis for MST, MSTS8 and MSTS30 at 5 K, respectively. The first harmonic dominates at low fields for all samples, whereas MSTS30 develops a rapidly increasing third-harmonic component A$_3$ with increasing field, becoming comparable to or larger than A$_1$ at low temperatures, as can be seen from Fig.~\ref{fig4}(c)-(f). The enhancement of A$_3$ occurs in the same temperature and field regime where the IPAHE is strongest, establishing a direct correlation between the two responses.

\begin{figure}
  \includegraphics[width=0.8\linewidth]{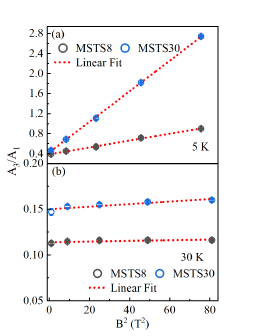}
  \caption{(a) , (b) Magnetic-field dependence of the ratio $\mathrm{A_3/A_1}$ for MSTS8 and MSTS30 at 5 K and 30 K, respectively. The data are plotted as a function of B$^2$, and the dotted red lines represent linear fits. Here, A$_1$ and A$_3$ denote the amplitudes of the first- and third-harmonic components extracted from the FFT analysis of the angular Hall resistivity.}
  \label{fig5}
\end{figure}

Figure~\ref{fig5}(a),(b) shows the ratio $\mathrm{A_3/A_1}$ as a function of magnetic field for both MSTS8 and MSTS30 at 5 K and 30 K, respectively. The data are plotted as a function of B$^2$, and the dotted red lines represent linear fits. Since $\mathrm{A_1}$ represents the conventional first-harmonic Hall response and $\mathrm{A_3}$ characterizes the symmetry-driven third-harmonic contribution, the ratio $\mathrm{A_3/A_1}$ provides a direct measure of the relative evolution of the higher-order transverse transport. At 5 K, where the IPAHE is most pronounced, $\mathrm{A_3/A_1}$ increases rapidly with magnetic field for both samples and follows a linear behaviour. Such behavior is consistent with the expectation that the third-harmonic Hall response originates from a higher-order transverse transport channel that grows more rapidly with magnetic field ($\propto \mathrm{B^3}$) than the conventional first-harmonic contribution ($\propto \mathrm{B}$). In MSTS30, $\mathrm{A_3/A_1}$ increases more rapidly and exceeds unity at high fields, indicating the dominance of third harmonic contribution over the first harmonic, which becomes almost negligible at higher field values (see Appendix\hyperref[appendixC]{C} for more details). In contrast, at 30 K, where the IPAHE is strongly suppressed, the field dependence of $\mathrm{A_3/A_1}$ becomes nearly field-independent. The weak variation of $\mathrm{A_3/A_1}$ indicates that the Hall response is dominated by the conventional first-harmonic component, while the higher-order symmetry-governed channel contributes only as a small correction.

Combining the field-dependent Hall measurements, angular Hall analysis, and harmonic decomposition, it appears that anisotropy tuning drives the system from a predominantly conventional Hall regime toward a symmetry-controlled nonlinear transport regime. The simultaneous enhancement of the IPAHE and the third harmonic, together with the nearly quadratic field dependence of $\mathrm{A_3/A_1}$ at low temperatures, provides compelling evidence suggesting that both phenomena originate from the same anisotropy-driven transport mechanism. Reducing the magnetic anisotropy strengthens the coupling between the in-plane magnetization and the trigonal crystal potential, allowing higher-order Hall channels to become experimentally observable (see Appendix\hyperref[appendixD]{D} for more details).

{\it Conclusion:---}
In summary, we have demonstrated the emergence of an in-plane anomalous Hall effect and a pronounced third-harmonic Hall response in an easy-plane trigonal magnetic system through anisotropy tuning. Se substitution progressively reduces the magnetic anisotropy, enhances the in-plane magnetic response, and promotes the development of a finite IPAHE. Angular-dependent Hall measurements reveal a clear departure from conventional first-harmonic behavior and the emergence of a strong $\mathrm{\sin(3\theta)}$ contribution, reflecting the underlying threefold rotational symmetry of the crystal lattice. Harmonic decomposition and FFT analysis show a systematic transfer of spectral weight from the first to the third harmonic, while the field dependence of $\mathrm{A_3/A_1}$ establishes the growing dominance of the nonlinear Hall channel at low temperatures and high magnetic fields. These results reveal a direct interplay between magnetic anisotropy, crystal symmetry, and transverse transport, and demonstrate that anisotropy tuning provides an effective route for engineering higher-order Hall phenomena in low-symmetry magnetic materials.

{\it Acknowledgements:---}
The authors acknowledge IIT Kanpur and the Department of Science and Technology, India, [Order No. DST/NM/TUE/QM-06/2019 (G)] for financial support. A.D. thanks the PMRF for financial support.

{\it Conflict of Interest:---}
The authors declare no conflict of interest.
\section{Appendix A: Preliminary Charecterization}\label{appendixA}
{\it Preparation:---}
Single crystals of Mn$_3$Si$_2$Te$_6$ (MST) have been prepared using the chemical vapour transport (CVT) method with iodine as the transport agent \cite{PhysRevB.111.174419}. The initial precursors, Mn (99.95$\%$, Alfa Aesar), Si (99.999$\%$, Alfa Aesar), and Te (99.99$\%$, Alfa Aesar) powders, are mixed in the stoichiometric ratio of 3:2:6. The growth process utilizes a TG3-1200 gradient three-zone tube furnace with the hot end kept at 800\degree C and the cold end maintained at 750\degree C. The ampoule is kept in the furnace for 20 days, and after that, it is cooled to room temperature and taken out. Mn$_3$Si$_2$(Te$_{1-x}$Se$_x$)$_6$ (MSTS) single crystals were also grown using the chemical vapour transport (CVT) method with iodine as the transport agent \cite{PhysRevB.111.174419}. Now, the Mn (99.95$\%$, Alfa Aesar), Si (99.999$\%$, Alfa Aesar), Te (99.99$\%$, Alfa Aesar) and Se (99.999$\%$, Alfa Aesar) powders, are mixed in the stoichiometric ratio and kept in the furnace for 17 days with the hot end at 750\degree C and the cold end at 700\degree C. Upon cooling to room temperature, we successfully obtained the crystals.

\begin{figure}
  \includegraphics[width=\linewidth]{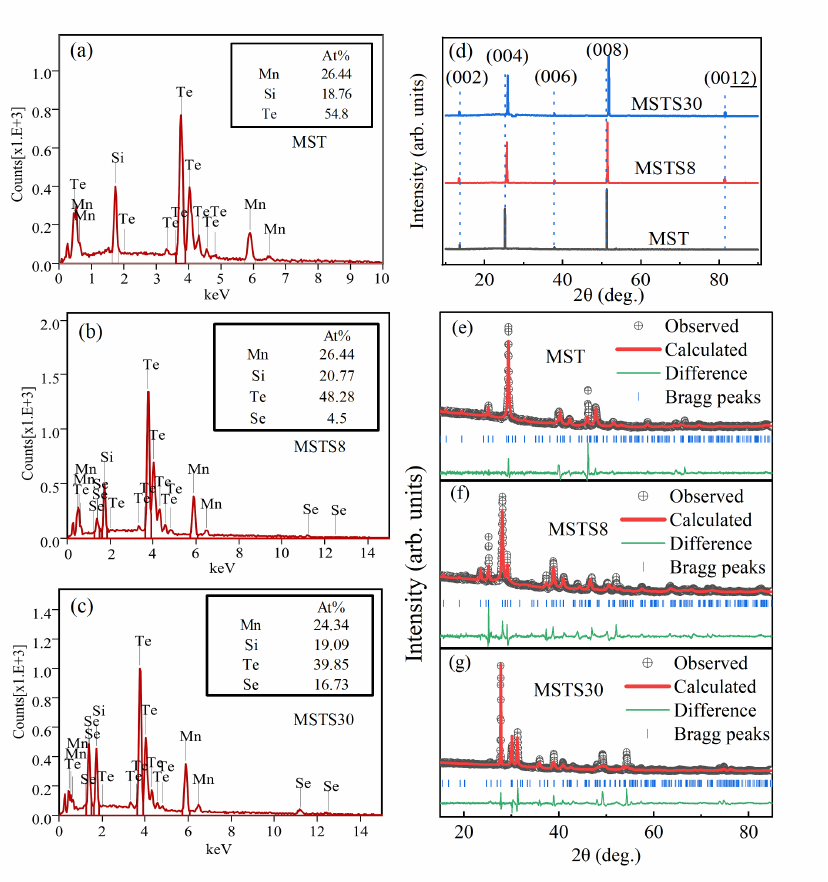}
  \caption{(a) - (c) Energy dispersive spectroscopy data of Mn$_3$Si$_2$(Te$_{1-x}$Se$_x$)$_6$ (MSTS$x$) single crystals for different values of $x$. (d) Single crystal XRD Data of Mn$_3$Si$_2$(Te$_{1-x}$Se$_x$)$_6$ (MSTS$x$) single crystals for different values of $x$. (e) - (g) Powder XRD Data of Mn$_3$Si$_2$(Te$_{1-x}$Se$_x$)$_6$ single crystals for different values of $x$.}
  \label{smfig1}
\end{figure}

{\it Charecterization:---}
Energy-dispersive spectroscopy (EDS) and X-ray diffraction (XRD) measurements are performed to determine the crystallinity as well as the elemental composition of both the substituted and unsubstituted samples. SM-Fig.~\ref{smfig1}(a)-(c) shows the EDS data of Mn$_3$Si$_2$(Te$_{1-x}$Se$_x$)$_6$ (MSTS$x$) single crystals for different values of $x$. It shows that the obtained stoichiometry is very close to the desired ones. The crystallinity of the sample is confirmed by single-crystal XRD as shown in SM-Fig.~\ref{smfig1}(d). Powder XRD (pXRD) measurement is also carried out using a PANalytical Empyrean diffractometer confirming the purity of the grown crystals and confirming that no additional phases are formed upon Se substitution. SM-Fig.~\ref{smfig1}(e)-(g) shows the pXRD data of MSTS$x$ single crystals for different values of $x$. The pXRD plots have been profile-fitted with the space group P$\Bar{3}$1c, which is the original space group of MST. The good quality of fit confirms that the trigonal crystal structure is preserved upon Se substitution. 

\section{Appendix B: Fitting of Hall resistivity}\label{appendixB}
\begin{figure}
  \includegraphics[width=\linewidth]{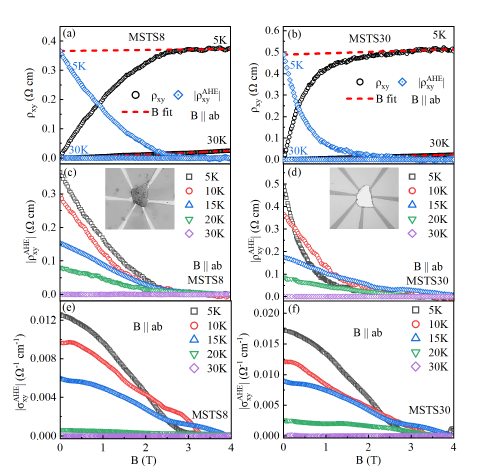}
  \caption{Field-dependent Hall resistivity, $\mathrm{\rho_{xy}}$, and extracted absolute in-plane anomalous Hall resistivity, $|\mathrm{\rho_{xy}^{AHE}}|$, for (a) MSTS8 and (b) MSTS30, respectively, at 5 and 30 K. The dashed red curves denote the corresponding B-dependent linear fits. (c), (d) The absolute in-plane anomalous Hall resistivity, $|\mathrm{\rho_{xy}^{AHE}}|$, and (e), (f) conductivity, $|\mathrm{\sigma_{xy}^{AHE}}|$, for MSTS8 and MSTS30, respectively, as a function of field for different temperatures. Inset (c),(d): Scanning electron microscopy image (SEM) of MSTS8 and MSTS30 nanoflake devices, respectively.}
  \label{smfig2}
\end{figure}

To eliminate the contribution of the ordinary Hall effect, we fit the Hall resistivity data of MSTS8 and MSTS30 at high fields and extrapolate them at lower field values. Subtracting the extrapolated fit with actual resistivity data gives the remnant in-plane anomalous Hall resistivity. SM-Fig.~\ref{smfig2}(a),(d) shows the field-dependent Hall resistivity, $\mathrm{\rho_{xy}}$, and extracted absolute anomalous Hall resistivity, $|\mathrm{\rho_{xy}^{AHE}}|$, for MSTS8 and MSTS30, respectively, at 5 and 30 K. The dashed red curves denote the linear B-fit, which is fitted at high field and extrapolated to low fields. It is evident from the figures that the extracted absolute anomalous Hall resistivity, $|\mathrm{\rho_{xy}^{AHE}}|$, shows a finite value at 5 K for both MSTS8 and MSTS30. However, at 30 K, $|\mathrm{\rho_{xy}^{AHE}}|$ is negligible for both the Se substituted samples. The temperature dependence of absolute in-plane anomalous Hall resistivity, $|\mathrm{\rho_{xy}^{AHE}}|$, and conductivity, $|\mathrm{\sigma_{xy}^{AHE}}|$, respectively, are shown in SM-Fig.~\ref{smfig2}(b)-(f) for both MSTS8 and MSTS30. It is clear that as the temperature is increased, the anomalous conductivity decreases and becomes negligible at 30 K. The inset of SM-Fig.~\ref{smfig2}(b),(e) shows the scanning electron microscopy (SEM) image of the nanoflake device of MSTS8 and MSTS30, respectively.
\begin{figure*}[htp]
  \includegraphics[width=0.9\linewidth]{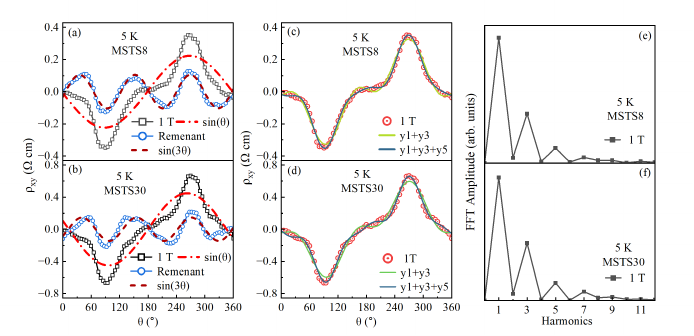}
  \caption{\textbf{Harmonic decomposition of the angular Hall resistivity at 5~K and $\mathbf{B=1}$~T.} Angular dependence of the Hall resistivity $\rho_{xy}(\theta)$ for (a) MSTS8 and (b) MSTS30. The black symbols represent the measured angular Hall resistivity. The dashed red curves represent the first-harmonic $\mathrm{\sin(\theta)}$ contribution, whereas the dashed brown curves represent the third-harmonic $\mathrm{\sin(3\theta)}$ contribution. The blue symbols represent the remnant angular Hall resistivity after eliminating the contribution of the first-harmonic $\mathrm{\sin(\theta)}$ from the measured data. (c,d) Reconstruction of the measured angular Hall resistivity for MSTS8 and MSTS30, respectively, using successive harmonic components, $\mathrm{A_1\sin\theta+A_3\sin(3\theta)}$ and $\mathrm{A_1\sin\theta+A_3\sin(3\theta)+A_5\sin(5\theta)}$. Here, yn corresponds to $\mathrm{\sin(n\theta)}$ (the nth order harmonic). (e,f) Fast Fourier transform (FFT) spectra of the angular Hall resistivity for MSTS8 and MSTS30, respectively, showing the amplitudes of the odd harmonic components.}
  \label{smfig3}
\end{figure*}
\section{Appendix C: Fitting and fast Fourier transform of angular Hall resistivity}\label{appendixC}
Since the angular Hall resistivity $\mathrm{\rho_{xy}}$ of both MSTS8 and MSTS30 show clear deviation from a purely sinusoidal angular dependence, we perform the harmonic decomposition of the angular Hall response. The conventional first-harmonic contribution, $\mathrm{\sin(\theta)}$, captures the overall variation of the Hall response but does not reproduce all the features of the experimental angular dependence, particularly around the extrema. It is clear from SM-Fig.~\ref{smfig3}(a),(b) that $\mathrm{\rho_{xy}}$ cannot be fitted only with $\mathrm{\sin(\theta)}$. Eliminating the contribution of $\mathrm{\sin(\theta)}$ from $\mathrm{\rho_{xy}}$ gives a remnant part for both MSTS8 and MSTS30, as is evident from SM-Fig.~\ref{smfig3}(a),(b) respectively, which gets fitted upon including the contribution of higher harmonics, i.e., $\mathrm{\sin(3\theta)}$. To ascertain the contribution of higher harmonics quantitatively, the experimental angular Hall data were reconstructed progressively using 
\begin{eqnarray}
      \mathrm{\rho_{xy}} &=& \mathrm{y1+y3} \label{1} \\
    \mathrm{\rho_{xy}} &=& \mathrm{y1+y3+y5} \label{2}
\end{eqnarray}
Here, y1, y3, y5 corresponds to the first, third and fifth harmonic [$\mathrm{\sin(\theta)}$, $\mathrm{\sin(3\theta)}$ and $\mathrm{\sin(5\theta)}$] respectively. From SM-Fig.~\ref{smfig3}(c),(d) we can see that the first and third harmonic reconstruction already reproduces the principal features of the measured angular Hall response. Inclusion of the fifth harmonic provides only a further refinement of the reconstruction, indicating that the dominant deviation from the conventional first-harmonic response originates from the third harmonic. SM-Fig.~\ref{smfig3}(e),(f) shows the corresponding Fast Fourier transform (FFT) spectra of the angular Hall resistivity for MSTS8 and MSTS30, respectively. We can see that the spectral weight is concentrated predominantly in the first and third harmonics, with progressively negligible contributions from higher odd harmonics. The successful reconstruction of the measured angular response provides direct experimental support for the emergence of a higher-order Hall channel upon reduction of the magnetic anisotropy. The FFT spectra amplitude supplements the above observation.

\begin{figure*}
  \includegraphics[width=0.8\linewidth]{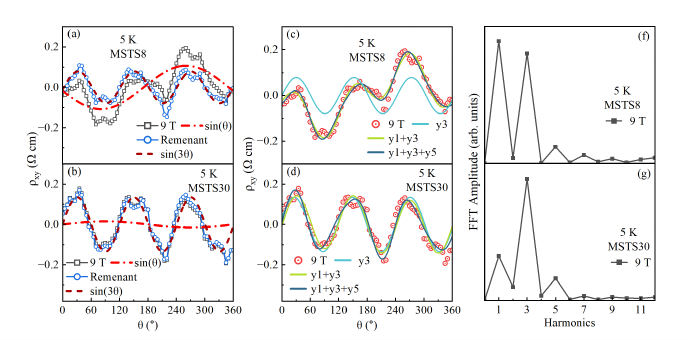}
  \caption{\textbf{Harmonic decomposition of the angular Hall resistivity at 5~K and $\mathbf{B=9}$~T.} Angular dependence of the Hall resistivity $\rho_{xy}(\theta)$ for (a) MSTS8 and (b) MSTS30. The black symbols represent the measured angular Hall resistivity. The dashed red curves represent the first-harmonic $\mathrm{\sin(\theta)}$ contribution, whereas the dashed brown curves represent the third-harmonic $\mathrm{\sin(3\theta)}$ contribution. The blue symbols represent the remnant angular Hall resistivity after eliminating the contribution of the first-harmonic $\mathrm{\sin(\theta)}$ from the measured data. (c,d) Reconstruction of the measured angular Hall resistivity for MSTS8 and MSTS30, respectively, using successive harmonic components, $\mathrm{A_1\sin\theta+A_3\sin(3\theta)}$ and $\mathrm{A_1\sin\theta+A_3\sin(3\theta)+A_5\sin(5\theta)}$. Here, yn corresponds to $\mathrm{\sin(n\theta)}$ (the nth order harmonic). (e,f) Fast Fourier transform (FFT) spectra of the angular Hall resistivity for MSTS8 and MSTS30, respectively, showing the amplitudes of the odd harmonic components.}
  \label{smfig4}
\end{figure*}

To further examine the evolution of the harmonic response with magnetic field, a similar analysis was performed at 9 T. At higher field, the angular Hall resistivity exhibits a pronounced departure from the simple first harmonic, $\mathrm{\sin(\theta)}$, dependence for both MSTS8 and MSTS30. The deviation is particularly evident in MSTS30, where the angular Hall response develops a clear threefold modulation. The corresponding contribution of the $\mathrm{\sin(3\theta)}$ component thus becomes substantially more important at high field. As in the 1 T analysis, the measured angular Hall resistivity was first fitted with the conventional first-harmonic contribution, $\mathrm{\sin(\theta)}$, and the remnant with $\mathrm{\sin(3\theta)}$. From SM-Fig.~\ref{smfig4}(a),(b) we can see that at 9 T, the remnant curve almost follows the original $\mathrm{\rho_{xy}}$ data even after eliminating the contribution of $\mathrm{\sin(\theta)}$. This is more pronounced for MSTS30, where the original $\mathrm{\rho_{xy}}$ data and the remnant curve almost superimpose upon each other. This highlights the fact that as the field is increased along with increasing Se concentration, the higher harmonic contribution, $\mathrm{\sin(3\theta)}$, becomes much more dominant than the first harmonics. Again, the experimental angular Hall data were reconstructed progressively using equation~\ref{1} and equation~\ref{2}. However, since at high magnetic fields, the higher harmonics show a more dominant contribution, we additionally try to reconstruct $\mathrm{\rho_{xy}}$ using only $\mathrm{\sin(3\theta)}$, i.e., $\mathrm{\rho_{xy}} = \mathrm{y3}$. SM-Fig.~\ref{smfig4}(c),(d) captures the reconstructed curves, and we can see that for MSTS8, $\mathrm{\sin(3\theta)}$ alone cannot completely reconstruct the obtained angular Hall data. The first harmonic contribution of $\mathrm{\sin(\theta)}$ is still plays a significant role for successful reconstruction of the obtained experimental data. Again, the inclusion of the fifth harmonic provides only a further refinement of the reconstruction. However, for MSTS30, the third harmonic contribution $\mathrm{\sin(3\theta)}$ alone reproduces the principal features of the measured angular Hall response. Along with the fifth harmonic, the conventional first harmonic term also acts only as a refining parameter of the reconstruction. The corresponding FFT spectra of the angular Hall resistivity for MSTS8 and MSTS30 are shown in SM-Fig.~\ref{smfig4}(e),(f), respectively. For MSTS8, the spectral weight is governed predominantly by the first and third harmonics, with an increased spectral amplitude for the third harmonic, almost equal in magnitude to the first harmonic. For MSTS30, the spectral weight is primarily governed by the third harmonic, which acts as the dominating term over the almost negligible first harmonic term. The progressive transfer of spectral weight toward the third harmonic upon increasing the magnetic field is therefore particularly pronounced in the more strongly anisotropy-tuned MSTS30 sample. 

Comparison with the corresponding 1 T data further demonstrates that the third-harmonic contribution is not simply a fixed geometrical feature of the angular measurement, but evolves systematically with magnetic field. Thus, the enhanced third-harmonic spectral weight at 9 T is consistent with the faster field growth of the nonlinear Hall channel relative to the first-harmonic contribution. This field evolution is consistent with the $\mathrm{A_3/A_1 \propto \mathrm{B^2}}$ behavior discussed in the main text and provides an independent visualization of the same harmonic evolution obtained from the field-dependent FFT analysis. The stronger high-field third-harmonic response in MSTS30 is also consistent with its reduced magnetic anisotropy and enhanced in-plane magnetization. As discussed in the main text, anisotropy tuning increases the in-plane magnetic response and strengthens the symmetry-allowed transverse transport channels. The observation of a pronounced third harmonic at high field, together with its enhancement upon Se substitution, therefore supports the connection between the anisotropy-tuned in-plane Hall response and the higher-order Hall channel permitted by the trigonal symmetry. The successful reconstruction of the measured angular dependence using the extracted harmonic components at both 1 T and 9 T further confirms that the third-harmonic contribution is a robust feature of the experimental Hall response rather than an artifact of the Fourier analysis.
\section{Appendix D: Symmetry-constrained origin of the in-plane anomalous Hall response and higher-order harmonics}\label{appendixD}
The anomalous Hall effect is conventionally described by an antisymmetric transverse conductivity. In the simplest description of a ferromagnet, the anomalous Hall conductivity is proportional to the magnetization through
\begin{eqnarray}
    \mathrm{\sigma^A_{ij}} &\propto& \mathrm{\epsilon_{ijk}M_k}
\end{eqnarray}
where $\mathrm{\epsilon_{ijk}}$ is the Levi-Civita tensor. For a current applied along the x direction, the conventional Hall conductivity is given as $\mathrm{\sigma^A_{xy} \propto \mathrm{M_z}}$. This explains why the conventional anomalous Hall effect is generally strongest when the magnetization has an out-of-plane component. Consequently, the conventional anomalous Hall response is expected to become small when the magnetization is confined to the basal plane as is the case for MST and MSTS$x$, which are easy-plane ferrimagnets with the ab plane as the magnetic easy plane. The magnetic anisotropy progressively decreases with Se substitution, accompanied by an enhanced in-plane saturation magnetization M$\mathrm{_{ab}}$. The observation of a finite Hall response for B $\|$ ab indicates that the Hall conductivity cannot be described solely by the conventional $\mathrm{M_z}$-dependent term. Previous reports also show the existence of a finite in-plane anomalous Hall response in monolayer MST \cite{PhysRevB.109.155153}. To describe this contribution, the Hall response may be expanded in terms of the magnetic field and magnetic order parameter. 

We know from Ohm's law $\mathrm{E_j = \rho_{ij}J_i}$, where $\mathrm{\rho_{ij} = R_{ijk}B_k}$ and i denotes the direction of applied current, j the direction of Hall voltage, and k the direction of applied magnetic field. Now, for a current applied along x, the transverse electric field along y can be written schematically as
\begin{eqnarray}
    \mathrm{E_y} &=& \mathrm{R_{xyz}J_x B_z + R_{xyx}J_x B_x} + \cdots .
\end{eqnarray}
The first term represents the conventional Hall response generated by the out-of-plane field component $\mathrm{B_z}$, whereas the second term represents an in-plane Hall channel. In particular, $\mathrm{E_y = R_{xyx}J_x B_x}$ defines the in-plane Hall coefficient $\mathrm{R_{xyx}}$. A finite $\mathrm{R_{xyx}}$ is not allowed by arbitrary crystal symmetries or systems with high crystal symmetries. Its existence requires that the magnetic point-group symmetry of the ordered state permits the coupling between the in-plane magnetic field and the transverse electric field. Absence of certain crystal symmetries and presence of C$_1$ or C$_3$ rotational symmetry along the z-axis is a necessary requirement for $\mathrm{R_{xyx}}$ to become finite \cite{PhysRevResearch.5.023138}, providing a symmetry-allowed route to the in-plane anomalous Hall effect (IPAHE). The connection between the magnetic state and the in-plane Hall response can be expressed phenomenologically by expanding the linear in-plane Hall coefficient in powers of the in-plane magnetic order parameter. For the easy-plane ferrimagnetic state considered here,
\begin{eqnarray}
   \mathrm{R_{xyx}} &=&
\mathrm{\lambda_1 M_{ab}
+
\lambda_3 M_{ab}^{3}
+\cdots}, 
\end{eqnarray}
where $\mathrm{M_{ab}}$ is the in-plane magnetization and $\lambda_1$ and $\lambda_3$ are material-dependent coefficients determined by the spin-orbit-coupled electronic structure and the crystal and magnetic symmetries. In the regime where the linear term dominates, $\mathrm{R_{xyx}}=\mathrm{\lambda_1 M_{ab}}$. Thus, as Se substitution reduces the magnetic anisotropy and increases the in-plane saturation magnetization, $\mathrm{M_{ab}}$ becomes progressively larger and the symmetry-allowed in-plane Hall coefficient  $\mathrm{R_{xyx}}$ is expected to increase, giving rise to the observed enhancement of the IPAHE.

The crystal structure of MST and MSTS$x$ possesses a threefold rotational symmetry about the c axis. This imposes constraints on the allowed angular dependence of the Hall response. In addition to the linear in-plane coupling described by $\mathrm{R_{xyx}}$, higher-order terms involving the in-plane magnetic-field components can be symmetry allowed. The cubic combinations of the in-plane field components have the form 
\begin{eqnarray}
    B_x^3-3B_xB_y^2 , \\
    B_y^3-3B_x^2B_y.
\end{eqnarray}
For an in-plane magnetic field written as $\mathrm{B_x=B_{ab}\cos\phi}$ and $B_y=B_{ab}\sin\phi$, on substituting becomes 
\begin{eqnarray}
    \mathrm{B_x^3-3B_{x}B_y^2} = \mathrm{B_{ab}^3\cos3\phi}, \\
    \mathrm{B_y^3-3B_x^2B_y} = \mathrm{-B_{ab}^3\sin3\phi}.
\end{eqnarray}
Thus, the threefold crystal symmetry permits a cubic magnetic-field contribution with a characteristic third-harmonic angular dependence. The Hall response can consequently be expressed schematically as 
\begin{eqnarray}
    \mathrm{\rho_{xy}^A} = \mathrm{\rho_{xy}^{(1)}} + \mathrm{\rho_{xy}^{(3)}{}} + \cdots , 
\end{eqnarray}
where the first term denotes the first order in-plane Hall channel contribution $\mathrm{\rho_{xy}^{(1)} \sim \mathrm{R_{xyx}J_x B_x}}$ and the second term refers to the leading symmetry-related nonlinear contribution $\mathrm{\rho_{xy}^{(1)} \sim \mathrm{C_3 B^3 sin(3\phi)}}$. Thus, the IPAHE and the higher harmonic  response have a common symmetry origin but represent different orders of the transverse response. When the magnetic field is rotated in the plane containing the c axis and the current direction, $\mathrm{B_z = B cos\theta}$ and $\mathrm{B_x = B sin\theta}$. The measured Hall resistivity can be phenomenologically decomposed as 
\begin{eqnarray}
    \mathrm{\rho_{xy}(\theta , B)} = \mathrm{R_{xyx} B sin\theta + R_{xyz} B cos\theta}.
\end{eqnarray}
For MSTS$x$, $\mathrm{R_{xyx} >> R_{xyz}}$ as $\mathrm{M_x >> M_z}$; thus, we can expand the measured angular Hall resistivity in odd powers of the magnetic field as
\begin{eqnarray}
    \mathrm{\rho_{xy}^{A}(\theta,B)} &=& \mathrm{\left(c_1B+c_1^{(3)}B^3+\cdots\right)\sin\theta} \\ && + \mathrm{\left(c_3B^3+c_3^{(5)}B^5+\cdots\right)\sin3\theta} + \cdots. \nonumber
\end{eqnarray}
Taking the lowest-order contribution to each channel as dominant we have $\mathrm{A_1^\rho(B) \simeq c_1 B}$ and $\mathrm{A_3^\rho(B) \simeq c_3 B^3}$. Consequently, 
\begin{eqnarray}
    \mathrm{\frac{A_3^\rho(B)}{A_1^\rho(B)}} = \mathrm{\frac{c_3 B^3}{c_1 B}} \\
    \mathrm{\frac{A_3^\rho(B)}{A_1^\rho(B)}} \propto \mathrm{B^2}.
\end{eqnarray}
The observed quadratic increase of $\mathrm{A_3/A_1}$ with field means that the third-harmonic contribution grows more rapidly with field than the first-harmonic contribution. Thus, the simultaneous enhancement of $\mathrm{M_{ab}}$, the IPAHE coefficient $\mathrm{R_{xyx}}$, and the third-harmonic amplitude with Se substitution, together with their common suppression at elevated temperature, provides a consistent phenomenological description of the anisotropy-tuned higher-order Hall response in the trigonal easy-plane ferrimagnet.

\bibliography{cite}

\end{document}